%% file: ConcCodeswNNs_v17_WCNC.tex
\documentclass[conference,10pt,letterpaper]{IEEEtran} 
\pdfoutput=1
\usepackage{amsmath,cite,amsfonts,amssymb,psfrag,amsthm,paralist,bm}
\usepackage{graphicx}
\usepackage{epstopdf}
\usepackage{rotating}
\usepackage{dsfont}
\usepackage{color}
\usepackage{multirow}

\usepackage{tikz_style}
\usepackage{standalone}

\usepackage{siunitx}

\usepackage[utf8]{inputenc}

\DeclareMathOperator*{\argmax}{arg\,max}

\begin{document}
\title{Concatenated Classic and Neural (CCN) Codes: ConcatenatedAE}
\IEEEoverridecommandlockouts
	
\author{%
    \IEEEauthorblockN{Onur~G\"unl\"u\textsuperscript{1,*}, Rick~Fritschek\textsuperscript{2,*}, and Rafael~F.~Schaefer\textsuperscript{3}
    }
    \IEEEauthorblockA{\textsuperscript{1}%
		Information Coding Division, Department of Electrical Engineering, Link{\"o}ping University, 
		onur.gunlu@liu.se
	}
    \IEEEauthorblockA{\textsuperscript{2}%
			Chair of Information Theory and Machine Learning, 
			TU Dresden,
			rick.fritschek@tu-dresden.de
	}
	\IEEEauthorblockA{\textsuperscript{3}%
			Chair of Information Theory and Machine Learning, 
			TU Dresden, 
			rafael.schaefer@tu-dresden.de
	}
}

\maketitle
\begingroup\renewcommand\thefootnote{*}\footnotetext{These authors contributed equally.}
\endgroup

%%%%%%%%%%%%%%%%%%%%%%%%%%%%%%%%%%%%%%%%%%%%%%%%%%%%%%%%%%%%%%%%%%%%%%%%%%%%%%%%%%%%%%%%%%%%%%%%%%%%%%%%%%%%%%%%%%%%
\begin{abstract}
    Small neural networks (NNs) used for error correction were shown to improve on classic channel codes and to address channel model changes. We extend the code dimension of any such structure by using the same NN under one-hot encoding multiple times, then serially-concatenated with an outer classic code. We design NNs with the same network parameters, where each Reed-Solomon codeword symbol is an input to a different NN. Significant improvements in block error probabilities for an additive Gaussian noise channel as compared to the small neural code are illustrated, as well as robustness to channel model changes.
\end{abstract}
%%%%%%%%%%%%%%%%%%%%%%%%%%%%%%%%%%%%%%%%%%%%%%%%%%%%%%%%%%%%%%%%%%%%%%%%%%%%%%%%%%%%%%%%%%%%%%%%%%%%%%%%%%%%%%%%%%%%
%
%\begin{IEEEkeywords}
%concatenated codes, over-complete autoencoder, error correction, concatenated classic and neural codes.
%\end{IEEEkeywords}
\IEEEpeerreviewmaketitle
%% %%%%%%%%%%%%%%%%%%%%%%%%%%%%%%%%%%%%%%%%%%%%%%%%%%%%%%%%%%%%%%%%%%%%
%% %%%%
%% %%%%               Introduction
%% %%%%
%% %%%%%%%%%%%%%%%%%%%%%%%%%%%%%%%%%%%%%%%%%%%%%%%%%%%%%%%%%%%%%%%%%%%%
\section{Introduction} \label{sec:intro}
Practical channel codes are necessary to reliably reconstruct messages that are transmitted through a noisy and/or nonlinear medium, such as air or optical fiber, with a low-complexity decoder that achieves small block error rates (BLER) and high code rates at specified blocklengths \cite{LinandCostello, Gallagerbook}. Capacity-achieving polar codes \cite{Arikan} are the first practical channel codes that are asymptotically optimal for binary-input memoryless symmetric channels. However, practical code design even for such channels, for which capacity-achieving and -approaching codes are known, requires careful adaptation of encoder-decoder pairs for different blocklengths, rates, channel parameters, and complexity constraints \cite{SCLPolar,UrbankeLDPCNotStable,ArikanPACArxiv}. Furthermore, a change in the channel model, especially to a non-standard one, can significantly deteriorate the reliability performance of a practical code designed for a given model, which necessitates the automation of the code design procedure. Thus, (deep) neural networks have been proposed in \cite{ProductAE,TurboAE,KOCodes,TenBrinkSerialTurboAE} as alternative encoder-decoder pairs used for error correction such that code design is automated with the aim of adapting to changes in the channel model and code parameters and of improving on classic channel codes in terms of BLER and complexity. However, due to the exponential increase in the number of codewords with the code dimension, large neural networks (NNs) need to be trained to design neural codes for large blocklengths and rates \cite{ProductAE}. We discuss a concatenation scheme used for classic channel codes, which is adapted below to inner neural codes to improve their performance and to make it possible to train them for larger blocklengths and rates.

Concatenating multiple classic codes to obtain a single code with a long blocklength that is decodable by multiple decoders, each of which is designed for short blocklengths, provides a low-complexity alternative to designing a classic code for long blocklengths directly \cite{ForneyConcatenatedReport}. The chain of the innermost encoder, noisy channel, and the corresponding innermost decoder can be equivalently seen as a \emph{superchannel}, to which the memorylessness property carries over from a memoryless channel. For the superchannel, one then designs an outer error-correcting code such that each output symbol of the outer code is an input to the innermost encoder. The outer code should then be nonbinary, so the most common outer codes used in practice are Reed-Solomon (RS) codes \cite{RSCodesFirst}, which are maximum-distance separable linear codes and defined over a large Galois field. Such a one-level concatenation achieves blocklengths and rates that are the products of the inner and outer codes' blocklengths and rates, respectively, where a coarse lower bound on the concatenated code's minimum distance can be obtained similarly \cite[pp.~740]{LinandCostello}. Thus, design of good codes for error correction with a large blocklength is possible by using such a concatenation scheme in which the decoding complexity increases only algebraically with the concatenated code's blocklength, whereas the BLER decreases exponentially with a smaller error exponent than the best achievable exponent for the given blocklength \cite{ForneyConcatenatedReport}. 

We propose to design classic channel codes for a given neural superchannel that is a chain of a neural encoder, communication channel, and neural decoder such that the blocklength and code dimension of the inner neural code can be extended linearly with the classic code's parameters, and the complexity increases algebraically. By training neural encoder-decoder pairs, which can be considered as over-complete autoencoders \cite{TurboAE} and each of which is the same NN with an input that is a different output symbol of the outer code, we design \emph{concatenated classic and neural (CCN)} codes. Such a concatenation allows to use one-hot encoding and categorical cross-entropy for training. Using an outer high-rate RS code with an errors-only decoder, we illustrate for an additive white Gaussian noise (AWGN) channel outstanding gains in terms of BLER as compared to the inner binary neural encoder-decoder pair. Using next an outer errors-and-erasures RS decoder by proposing a thresholding algorithm to define a neural decoder output symbol as erased, we illustrate further gains. Comparisons with the normal approximation for the finite blocklength performance \cite{Polyanskiy} are given. We also show that CCN codes with an intermediate block interleaver are robust to Rayleigh fast fading and bursty channels, following also because concatenation provides robustness against error bursts~\cite{ForneyConcatenatedReport}. Possible significants improvements to the code constructions and NN designs are listed.

%%%%%%%%%%%%%%%%%%%%%%%%%%%%%%%%%%%%%%%%%%%%%%%%%%%%%%%%%%%%%%%%%%%%%%%%%%%%%%%%%%%%%%%%%%
\emph{Notation}: Upper case letters, e.g. $X$, represent random variables and lower case letters, e.g. $x$, their realizations. A superscript denotes a sequence of variables, e.g. $\displaystyle X^n\!=\!X_1,X_2,\ldots, X_i,\ldots, X_n$, and a subscript, e.g. $i$, denotes the position of a variable in a string. Alternatively, boldface lower case letters, e.g. $\mathbf{x}$, represent vectors of random variable realizations with elements $\mathbf{x}_i$, and boldface upper case letters, e.g. $\mathbf{X}$, matrices of realizations. Calligraphic letters, e.g. $\displaystyle \mathcal{X}$, denote sets. $\mathbb{F}_q$ denotes a Galois field with $q$ elements, where $q$ is a prime power.   

%%%%%%%%%%%%%%%%%%%%%%%%%%%%%%%%%%%%%%%%%%%%%%%%%%%%%%%%%%%%%%%%%%%%%%%
\section{Problem Definition and Neural Code Design}
\label{sec:problemDef}
Consider the classic point-to-point channel model in which a message $u^k\in\{0,1\}^k$, selected uniformly at random, is encoded into a codeword $x^n\in\mathcal{X}^n$ under a block power constraint, i.e., $\sum_{i=1}^n |x_i|^2\leq nP$ for all codewords, that is sent through a noisy channel with output $y^n\in\mathcal{Y}^n$. The decoder estimates the transmitted codeword as $\widehat{x}^n\in\mathcal{X}^n$ that is mapped to the message $\widehat{u}^k\in\{0,1\}^k$. Denote the channel code rate of such an $(n,k)$ code as $R=k/n$. Define the bit error rate (BER) and block error rate (BLER) as
\begin{align}
	&\text{BER}\!=\!\frac{1}{k}\! \sum_{i=1}^k\!\Pr[U_i\!\neq\! \widehat{U}_i],\\
	&\text{BLER}=\Pr[U^k\neq \widehat{U}^k].
\end{align}

The main channel model we consider, without loss of essential generality (w.l.o.e.g.), is a real discrete-time AWGN channel such that $Y^n=X^n+Z^n$, in which the noise sequence $Z^n$ is independent of $X^n$ and is independent and identically distributed (i.i.d.) according to a Gaussian distribution with zero mean and variance $\sigma^2=N_0/2$, where $N_0$ is the noise power per positive-frequency. Suppose, w.l.o.e.g., $P=1$ and define the signal-to-noise ratio (SNR) as $\text{SNR}=1/N_0$, so we have $E_b/N_0 = 1/(N_0R)$, where $E_b$ is the average signal energy per information bit.

We next summarize the steps we use to design neural encoder-decoder pairs for error correction by using end-to-end learning, which are then concatenated with outer classic channel codes in Section~\ref{sec:CCNConstruction} below to make them practical.

%%%%%%%%%%%%%%%%%%%%%%%%%%%%%%%
\subsection{Neural Channel Encoder and Decoder Design}
Consider binary neural encoder-decoder pairs with $k_1$ input bits in which the neural encoder first bijectively maps an input sequence $\tilde{u}^{k_1}$ to a one-hot-encoded binary vector $\mathbf{s}\in \mathbb{F}_2^{2^{k_1}}$, i.e., only the $j$-th bit $\mathbf{s}_j$ of $\mathbf{s}$ corresponds to the sent sequence, where $j\in \{1,2,\ldots,2^{k_1}\}$ is the index of the message to be sent according to a total ordering of messages. The vector $\mathbf{s}$ is then mapped to an output symbol $\tilde{\mathbf{x}}\in\mathbb{R}^{n_1}$ by applying an encoding function $f_{\bm{\theta}}$, defining an $(n_1,k_1)$ neural channel code for $n_1\!\geq\! k_1\!>\!0$. The parametric encoding function consists of a sequence of affine maps $F_1, F_2, \ldots, F_L$ and non-linear activation functions $\sigma_1(\cdot), \sigma_2(\cdot), \ldots, \sigma_{L-1}(\cdot)$ such that
\begin{align}
    \mathbf{s} \mapsto F_L(\sigma_{L-1}( F_{L-1}(\sigma_{L-2}(\ldots \sigma_1 (F_1(\mathbf{s}))))))
\end{align}
where the affine map $F_\ell$ of the $\ell$-th layer for $\ell =1,2,\ldots, L$ comprises a weight matrix $\mathbf{W}_\ell$ and a bias vector $\mathbf{b}_\ell$. A commonly used non-linear activation function is the ReLU activation function, i.e., $\sigma_{\text{ReLU}}(\cdot)=\max\{0,\cdot\}$, which we use for the hidden layers. We collect all learnable weights and biases in a set of parameters, denoted as $\bm{\theta} = \{\mathbf{W}_1,\mathbf{b}_1,\ldots, \mathbf{W}_L, \mathbf{b}_L\}$. These weights can be optimized by supervised learning, i.e., by observing a dataset of $m$ input and output sample pairs $\{(\mathbf{s}_i,f(\mathbf{s}_i))\}_{i=1}^m$ and tuning the parameters $\bm{\theta}$ of the NN such that an empirical loss $\tfrac{1}{m}\sum_{i=1}^m l(f_{\bm{\theta}}(\mathbf{s}_i),f(\mathbf{s}_i))$ is minimized. In particular, we design an encoder with two hidden layers that use ReLU activation functions and with a linear output layer. All layers have the same number of nodes as the input size, except the last layer whose dimensionality is matched to the neural encoder output size. Furthermore, we enforce the power constraint by applying the normalization $\displaystyle \tilde{\tilde{\mathbf{x}}}=({\tilde{\mathbf{x}}-{\bm\mu}_{\tilde{\mathbf{x}}}})/{\sigma}_{\tilde{\mathbf{x}}}$, where ${\bm\mu}_{\tilde{\mathbf{x}}}=\mu_{\tilde{\mathbf{x}}}\mathbf{1}$ is the broadcasted mean and ${\sigma}^2_{\tilde{\mathbf{x}}}$ is the variance. Then, $\tilde{\tilde{\mathbf{x}}}$ is transmitted through a noisy channel. The neural decoder is a parametric decoding function $g_{\bm{\theta'}}(\cdot)$ that comprises two hidden layers and a linear output layer with softmax activation, where the softmax function transforms the last layer's output to a vector $\hat{\mathbf{s}}\in (0,1)^{2^{k_1}}$; see below for its definition. This vector can be interpreted as a vector of probabilities, in which the $\tilde{j}$-th element is the estimated probability that the $\tilde{j}$-th message was sent for $\tilde{j}\in\{1,2,\ldots,2^{k_1}\}$. The estimated message at the neural decoder is then $\argmax_{\tilde{j}}\, \hat{\mathbf{s}}_{\tilde{j}}$. An interpretation is that the NN outputs confidence scores and the estimated message is chosen to be the one with the highest score. To take advantage of the confidence scores that provide soft information, below we define a thresholding algorithm to define a block erasure as the NN decoder output by setting a necessary minimum confidence score for all decoded messages. This algorithm is later shown to be useful for applying an outer errors-and-erasures RS decoder.

We apply the categorical cross-entropy loss for our optimization process, i.e., $\displaystyle -\sum_{\tilde{j}=1}^{2^{k_1}}\mathbf{s}_{\tilde{j}} \log \hat{\mathbf{s}}_{\tilde{j}}=-\log \hat{\mathbf{s}}_j$, where $j$ is the sent message index and also the only non-zero bit index in $\mathbf{s}$. The loss is averaged over a given batch size $m$, resulting in the empirical loss $-\tfrac{1}{m}\sum_{i=1}^m\log (\hat{\mathbf{s}}_j)_i$ and used with a gradient-based optimization method to optimize the weights of the parametric neural functions.

\subsection{Single-label vs. Multi-label Approach}
We apply one-hot encoding to the NN inputs and pair them with a categorical cross-entropy loss for the system optimization (training). One can see this framework as a single-label approach since only one label is active at every step. It is generally not possible to use this approach for large neural code dimensions $k_1$ or blocklengths $n_1$, since the number of input nodes grows exponentially with $k_1$ and output nodes linearly with $n_1$. Thus, previous works apply a multi-label approach that uses binary encoding and a binary cross-entropy loss, where multiple labels (bits) can be active, i.e., set to one, simultaneously. However, using the multi-label approach has a few disadvantages. For instance, although the input layer complexity is reduced by the multi-label approach, the number of nodes required by the multi-label hidden layers is usually a constant multiple of the number required by the single-label hidden layers to achieve the same learning performance. Thus, the overall complexity for the multi-label approach is higher. Moreover, optimizing with the binary cross-entropy results in NN parameters that minimize an approximation of the reliability per bit, i.e., an approximate BER. However, error correcting code designs should aim to minimize the BLER, which equivalently minimizes the message decoding error. Using the categorical cross-entropy, we enforce optimizations on neural decoder output sequences, which are the non-binary symbols of the outer code, so that we design NN parameters by approximating the symbol error rate of the CCN code. Relating the number of symbol errors for the outer code to BLER for the CCN code, our code construction enforces to minimize the BLER; see Section~\ref{sec:CCNConstruction} below. Since CCN codes concatenate a long-blocklength outer code with a small neural encoder, one can use a categorical cross-entropy for the small inner NN, unlike previous constructions. Using soft information at the neural decoder outputs, i.e., confidence scores, further gains can be achieved, illustrated below by proposing a thresholding algorithm for CCN code symbol erasures.

\begin{figure*}[t]
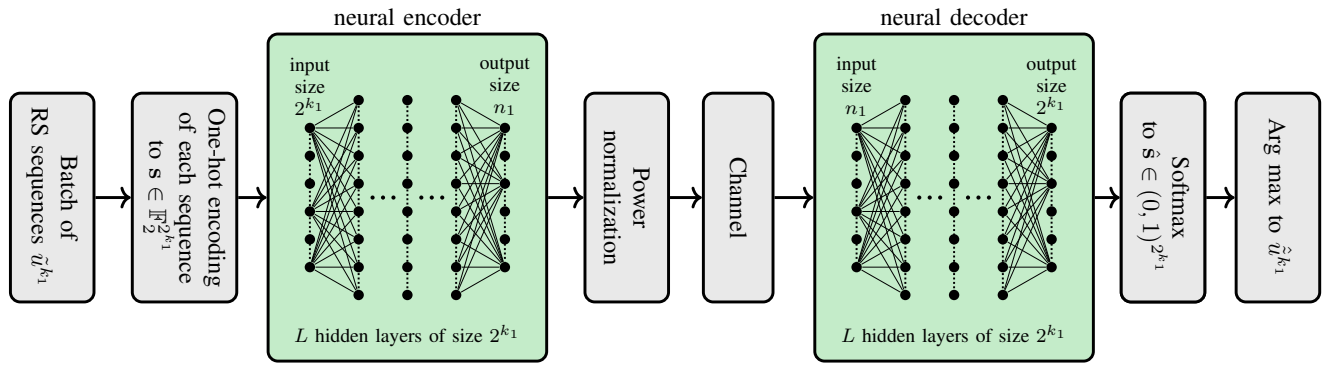

	\centering
	%\vspace{-0.15cm}
	\resizebox{\linewidth}{!}{
  \includestandalone[width=\textwidth]{Figure1}%     without .tex extension
  }
	%\vspace{-0.1cm}
	\caption{Inner neural encoder-decoder pair framework for CCN codes with an outer RS code and hard decoder decisions.}
	\label{fig:Neuralencdectrainingsteps}	
	%\vspace{-0.4cm}
\end{figure*}

%%%%%%%%%%%%%%%%%%%%%%%%%%%%%%%%%%%%%%%%%%%%%%%%%%%%%%%%%%%%%%%%%%%%%%%%
\section{CCN Code Constructions}\label{sec:CCNConstruction}
The code concatenation idea aims to reduce the total decoding complexity and delay for large blocklengths. To design concatenated codes, a binary inner code with a high error-correction capability and a small code dimension is generally used such that the remaining errors (and possibly also erasures) can be corrected by a high-rate non-binary outer code with a low-complexity decoder by treating bit sequences as its symbols \cite{ForneyConcatenatedReport}. Since there are various low-complexity algebraic methods to implement an RS decoder and since RS code symbols can be elements of any field $\mathbb{F}_q$, we concatenate outer $(n_2,k_2)$ RS codes with inner binary $(n_1,k_1)$ NN encoders, such that each RS codeword symbol, that is represented with $\log_2(q)=k_1$ bits, is an input to the same NN, i.e., the same inner NN encoder/decoder is used $n_2$ times. This concatenation is an $(n\!=\!n_2\cdot n_1,\;k\!=\!k_2\cdot\log_2(q))$ CCN code that is decoded by a neural decoder followed by a decoder for the outer RS code. A classic RS decoder is the errors-only decoder that can correct all error patterns with $e$ symbol errors such that $e\leq \lfloor(n_2-k_2)/2\rfloor$, and similarly an RS errors-and-erasures decoder can correct all error and erasure patterns with $e$ errors and $r$ erasures if $(2e+r)\leq (n_2-k_2)$ \cite[Section~7.7]{LinandCostello}, both of which are used for the CCN code constructions below.

RS codes protect against bursty errors caused by, e.g., the memory in the noisy channel, so it is advantageous to use outer RS codes especially when the noisy channel has a memory that is less than the inner neural code's blocklength \cite[Section~1.4]{ForneyConcatenatedReport}. Protection against burst errors and robustness to channel model changes can be improved by inserting an interleaver between the RS and neural encoders with a corresponding de-interleaver between neural and RS decoders. For all CCN code constructions below, we use a row-column block interleaver, in which $n_2$ RS codewords form an $n_2\times n_2$ matrix such that each row has the symbols of one codeword. The interleaver outputs are the elements of the matrix that are read out column-wise and given as inputs to the NN encoder. This interleaver allows to decode the symbols of a given RS codeword independently \cite[Section~1.4]{ForneyConcatenatedReport}.

CCN codes are different from the concatenation schemes in \cite{TenBrinkCodedModulationNNLDPC,NybridNeuralCodedModulation} in which an inner neural encoder is used only for constellation shaping and labeling, whereas our inner neural encoder is designed, w.l.o.e.g., as an error correcting code with real-valued outputs under a block power constraint. Furthermore, neural code concatenation schemes in \cite{ProductAE,TurboAE,TenBrinkSerialTurboAE} aim to design all component codes as NNs, whereas CCN codes aim to benefit from classic (non-binary) channel codes that are robust and have low-complexity decoders. Similarly, CCN codes reduce the required amount of channel simulations especially for high SNRs as compared to end-to-end neural code constructions. This simplification in the training follows since the error correction capability of the outer classic code already provides a coarse target symbol error probability for the small inner NN given a channel model, which seems to not depend strongly on the code rate or the target BLER for an AWGN channel \cite[Section~6.1.a]{ForneyConcatenatedReport}. 

We remark that the inner neural encoder and decoder of a CCN code can in principle be any neural code, including the ones proposed in \cite{ProductAE,TurboAE,TenBrinkSerialTurboAE,TenBrinkCodedModulationNNLDPC,NybridNeuralCodedModulation} by adapting our code concatenation scheme to the requirements of the given neural code (and modulation) constructions. Such combinations might allow to benefit from existing high-reliability neural constructions that can be designed only for small blocklengths.

%%%%%%%%%%%%%%%%%%%%%%%%%%%%%%%%%%%%%%%%%%%%%%%%%%%%%%%%
\section{Training and Experimental Results}
We consider a $(7,4)$ neural code as the reference code construction for our simulations to illustrate various effects of concatenation with an outer classic code. First, we aim to compare the reference neural code with a CCN code that has approximately the same code rate, i.e., $R_{\text{ref}}=4/7\approx 0.571$. Thus, we concatenate the outer $(255,223)$ RS code, that has symbols from $\mathbb{F}_{2^8}$, with an inner $(12,8)$ neural code such that CCN code rate is $(8\cdot 223)/(12\cdot 255)\approx 0.583$. The NN set-up applied to such a CCN code is depicted in Fig.~\ref{fig:Neuralencdectrainingsteps}, which includes the one-hot encoding of a batch of RS symbols that are then given as input to the neural encoder. Since the blocklength of the outer RS code is $n_2=255$, we choose a batch size of $m=n_2$ such that the neural encoder-decoder pairs are optimized over outer codewords. The row-column block interleaver, defined above, is applied before the training steps given in Fig.~\ref{fig:Neuralencdectrainingsteps} to increase the protection against burst errors and robustness to changes in the channel model. The de-interleaver is then applied before the RS decoding. 

Another motivation for concatenated codes is that by using a high-rate outer code, one can achieve larger blocklengths without decreasing the code rate significantly. Thus, we also consider a CCN code that comprises an outer $(15,11)$ RS code with symbols from $\mathbb{F}_{2^4}$ and the inner $(7,4)$ neural reference code such that concatenation decreases the rate from $R_{\text{ref}}\approx 0.571$ to $(4\cdot 11)/(7\cdot 15)\approx 0.419$, whereas the blocklength increases from $n_1=7$ to $n=15\cdot 7=105$. 

Concatenations with an outer code decreases the code rate and increases the blocklength simultaneously, which makes fair comparisons with other codes difficult. However, we illustrate that CCN codes with large code rates can be trained for large blocklengths, such as $n=255\cdot 12=3060$, whereas most neural channel codes in the literature have a blocklength of at most $256$ due to their high training complexity, which shows the eminent practicality of CCN codes. Furthermore, one can show that CCN codes perform worse than polar and LDPC codes, which are used in 5G new radio (NR), designed for binary-input AWGN (BI-AWGN) channels in terms of the BLER performance. However, we illustrate below that the gap to the normal approximation for BI-AWGN channels at finite blocklengths, calculated by using \cite{SpectreBIAWGNPaper}, is not too large for the BLER curves achieved by our CCN codes that are trained for and simulated in AWGN channels. This comparison is reasonable as we consider that neural codes bypass modulation, as mentioned also in \cite{KOCodes}. Moreover, we train and simulate CCN codes in a Rayleigh fading setting to illustrate concatenation gains, depicted in Fig.~\ref{fig:Rayleigh} below. To show that CCN codes are robust to changes or imprecision in the channel model, we also simulate CCN codes trained for AWGN channels in a bursty channel; see Fig.~\ref{fig:burst} below.

We train the inner neural encoder-decoder pairs for our CCN code by using the state-of-the-art end-to-end training methods with notable differences. For training, we use the Nadam optimizer \cite{Dozat2016Nadam} with a learning rate of \num{5e-4} and observe similar results with the plain stochastic gradient descent, the latter of which might be useful for generalizability but this study is outside the scope of this paper. We train with \num{e6} samples over $5$ epochs at $E_b/N_0=\{5,10,3\}$ dB for the AWGN, Rayleigh fast fading, and bursty channels, respectively, that are chosen to achieve a symbol error rate of approximately $10^{-2}$ for the inner neural codes. The standard autoencoder (AE) $(7,4)$ code is trained with \num{5e7} samples over $10$ epochs at the same $E_b/N_0$ training points as the corresponding CCN codes. We train the CCN code system as one, i.e., we do not split encoder and decoder trainings. Unlike some recent works, our decoder has the same complexity as our encoder, and we expect that the randomness that stems from the stochastic gradient-based optimization and the channel suffices to escape saddle points in the optimization process.

%%%%%%%%%%%%%%%%%%%%%%%%%%%%%%%
\subsection{Thresholding Algorithm to Implement RS Symbol Erasures}
To leverage further decoding capabilities of outer RS codes, we consider also an errors-and-erasures decoder that is a generalized minimum distance decoder \cite[Section~2.3]{ForneyConcatenatedReport} and that considers both symbol erasures and errors as the outputs of the neural decoders. Defining a neural decoder output sequence as an erased RS symbol requires a thresholding algorithm adapted to the parameters of the NN. Thus, we consider the softmax function $\sigma:\mathbb{R}^{2^{k_1}}\rightarrow (0,1)^{2^{k_1}}$ applied to the output ${\mathbf{z}}$ of the neural decoder's output layer to obtain the vector $\hat{\mathbf{s}}$, where $\displaystyle \sigma({\mathbf{z}})_{\tilde{j}}=e^{{\mathbf{z}}_{\tilde{j}}}\big/ \Big(\sum_{j=1}^{2^{k_1}} e^{{\mathbf{z}}_j}\Big)$ for all $\tilde{j}=1,2,\ldots,2^{k_1}$; see Fig.~\ref{fig:Neuralencdectrainingsteps}. Thus, the softmax function outputs are positive with a sum of one, so one can view $\hat{\mathbf{s}}$ as a probability mass function such that its $\tilde{j}$-th element is the probability that the $\tilde{j}$-th RS symbol's one-hot encoded representation was given as input to the corresponding neural encoder for $\tilde{j}=1,2,\ldots, 2^{k_1}$. Using this soft information, the decoder can infer a confidence level about the most likely RS symbol, i.e., $\bar{j}$-th RS symbol such that $\bar{j}=\argmax_{\tilde{j}}\, \hat{\mathbf{s}}_{\tilde{j}}$. In our simulations at a fixed $E_b/N_0$ that is chosen as a point close to the middle of the $E_b/N_0$ range of interest, we observe that defining the decoder output as an erased RS symbol if $\hat{\mathbf{s}}_{\bar{j}}\leq 0.5$ provides the best BLER results for AWGN channels in combination with an outer errors-and-erasures RS decoder. Thus, the threshold value we use below to define symbol erasures as the inner decoder output is $0.5$. The threshold can be further improved for each $E_b/N_0$ and channel model, which is left for future work.

\subsection{Experimental Results}
\begin{figure}[t]
    \centering
    %\vspace{-0.7cm}
    \includegraphics[width=1.06\linewidth,height=6cm]{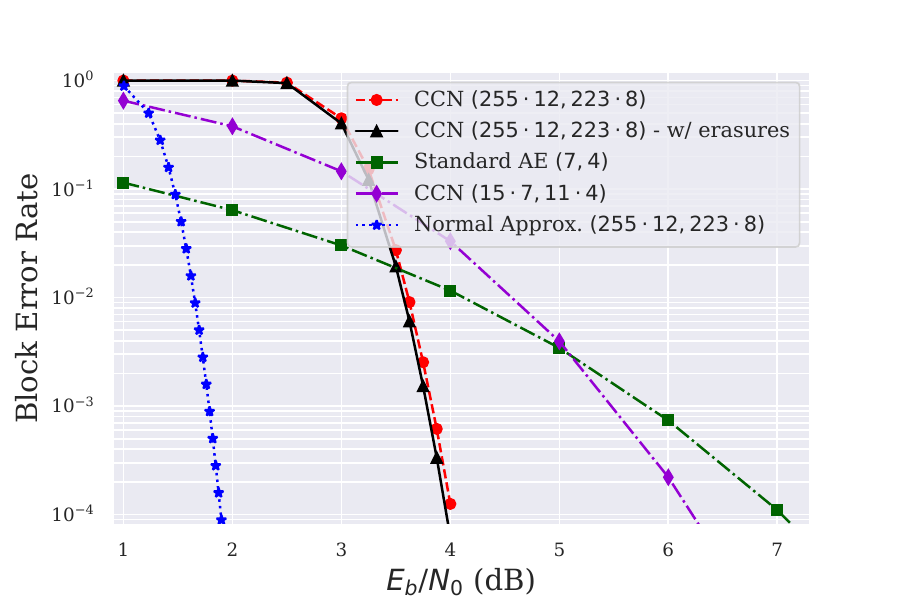}
    %\vspace{-0.7cm}
    \caption{BLER comparisons for the AWGN channel. $(255\cdot 12,223\cdot 8)$ CCN codes with an errors-only or errors-and-erasures RS decoder are compared with the normal approximation for given code parameters and with the standard AE $(7,4)$ reference code, which is compared also with a $(15\cdot 7,11\cdot 4)$ CCN code under errors-only decoding.}
    \label{fig:AWGN}
   % \vspace{-0.4cm}
\end{figure}

We consider a standard AE $(7,4)$ code as the reference code to illustrate the BLER gains from CCN codes as compared to the reference code. We first consider the AWGN channel to compare the reference code with CCN codes that have approximately the same code rate or that are constructed by concatenating the reference code with a high-rate outer code. Suppose either errors-only or errors-and-erasures RS decoders as the outer decoder of the CCN code to illustrate the effects of defining symbol erasures; see Fig.~\ref{fig:AWGN} for AWGN results of codes trained for this channel. The $E_b/N_0$ gain from a $(15\cdot 7, 11\cdot 4)$ CCN code as compared to the $(7,4)$ reference code at a BLER of $10^{-4}$ is approximately $0.7$~dB. Similarly, at the same BLER the $E_b/N_0$ gain from a $(255\cdot 12,223\cdot 8)$ CCN code is approximately $3.1$~dB, where the BLER improvements due to implementation of symbol erasures are negligible. Furthermore, the gap between the normal approximation for a $(255\cdot 12,223\cdot 8)$ code at BLER of $10^{-4}$ and the CCN codes designed with the same parameters is approximately $2.1$~dB, part of which is due to the performance degradation caused by concatenation.

%%%%%%%%%%%%%%%%%%%%%%%%%%%%%%%%%%%%%%%%%%%%%%%%%
%%%%%%%%%%%%%%%%%%%%%%%%%%%%%%%%%%%%%%%%%%%%%%%%%%%%

\begin{figure}
	\centering
	%\vspace{-0.6cm}
	\includegraphics[width=1.06\linewidth,height=6cm]{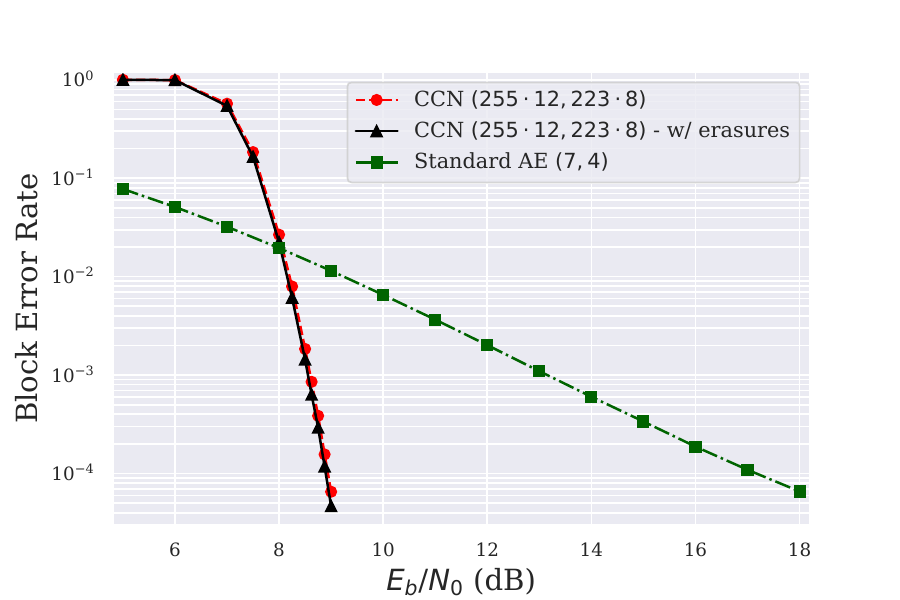}
	%\vspace{-0.3cm}
	\caption{BLER comparisons between $(255\cdot12,223\cdot 8)$ CCN codes and the standard AE $(7,4)$ reference code for the Rayleigh fast fading channel.}
	\label{fig:Rayleigh}
	%\vspace{-0.5cm}
\end{figure}

\begin{figure}
	\centering
	\includegraphics[width=1.06\linewidth,height=6cm]{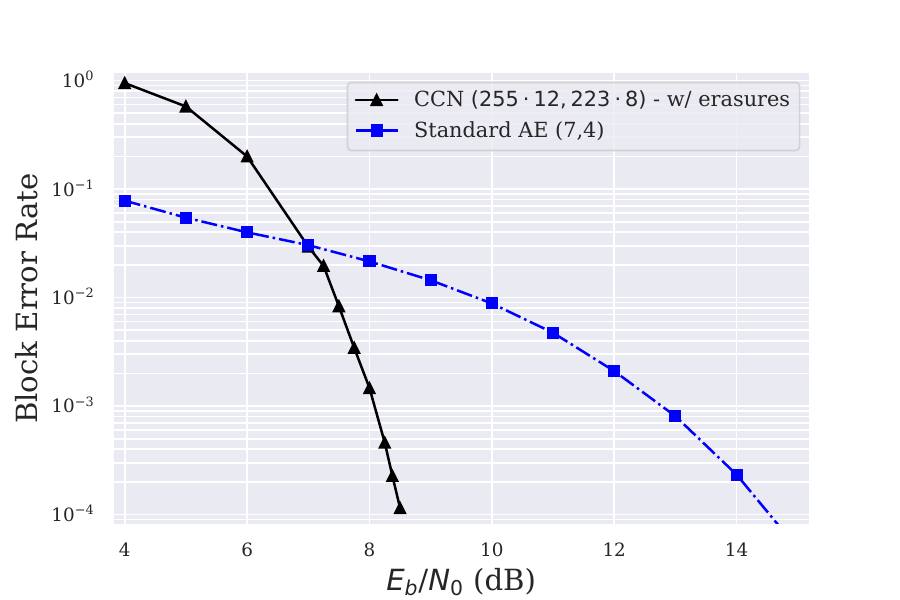}
	%\vspace{-0.3cm}
	\caption{BLER comparisons for testing robustness of an AWGN-trained $(255\cdot~\!\!\!12,223\cdot 8)$ CCN code by simulating it over a bursty channel and comparing its performance with the bursty channel performance of the standard AE $(7,4)$ reference code trained for the AWGN channel.}
	\label{fig:burst}
	%\vspace{-0.6cm}
\end{figure}

Suppose next a Rayleigh channel $Y_i=H_iX_i+N_i$ for $i=1,2,\ldots,n$, where $(H_i,X_i,N_i)$ are mutually independent, $X_i$ is the channel input under a block power constraint for all codewords $X^n$, $H_i$ is the Rayleigh channel coefficient, with $\mathbb{E}[H_i^2]=1$, and $N_i\sim \mathcal{N}(0,\sigma^2)$. Furthermore, we do not assume that the channel state information is available at the NNs. BLER results for the Rayleigh channel are depicted in Fig.~\ref{fig:Rayleigh} to compare the standard AE $(7,4)$ reference code with $(255\cdot 12,223\cdot 8)$ CCN codes, both of which are trained for the Rayleigh channel. We observe that CCN codes under errors-and-erasures decoding provide a significant $E_b/N_0$ gain of approximately $8.3$~dB at a BLER of $\num{e-4}$, which illustrates the robustness of CCN codes to fading channels. Moreover, the gap between the performances of errors-and-erasures decoding and errors-only decoding is less than $0.05$~dB. Now, consider a bursty channel modeled as $Y_i=X_i+N_i+W_i$, where $(X_i,N_i,W_i)$ are mutually independent, $N_i\sim\mathcal{N}(0,\sigma^2)$, and $W_i= C_iD_i$ such that $C_i\sim\mathcal{N}(0,2\sigma^2)$ and $D_i$ is a Bernoulli random variable with $\Pr[D_i=1]=p$, for which we choose $p=0.1$ as in \cite{KOCodes}. We simulate the AWGN-trained standard AE $(7,4)$ reference code and $(255\cdot 12,223\cdot 8)$ CCN code under errors-and-erasures decoding in the bursty channel and the BLER results are given in Fig.~\ref{fig:burst}. We observe that CCN codes are robust to the bursty channel and provide an $E_b/N_0$ gain of approximately $6.2$dB as compared to the reference code at a BLER of $\num{e-4}$.

%%%%%%%%%%%%%%%%%%%%%%%%%%%%%%%%%%%%%%%%%%%%%%%%%%%%
%%%%%%%%%%%%%%%%%%%%%%%%%%%%%%%%%%%%%%%%%%%%%%%%%%%%

%%%%%%%%%%%%%%%%%%%%%%%%%%%%%%%%%%%%%%%%%%%%%%%%%%%%
%%%%%%%%%%%%%%%%%%%%%%%%%%%%%%%%%%%%%%%%%%%%%%%%%%%%%

For all results, we test the CCN codes with a total of $\num{127500}$ RS codewords with \numrange{1}{80} noise samples each, resulting in a total sample size between \num{5e5} at the lowest $E_b/N_0$ and approximately $10^7$ at the highest $E_b/N_0$, respectively. The reference code is tested with a sample size of \num{5e7} at each $E_b/N_0$. Furthermore, BLER performance of CCN codes can be significantly improved, e.g., by applying list decoding to the outer (folded) RS code via the reliability information obtained from the neural decoder outputs that can also be a list and by optimizing the interleaver used. Similarly, an iterative decoding method can be used to improve the BLER performance. Moreover, our CCN code constructions that use cross-entropy loss for optimizations can be extended for the estimated-mutual-information loss. Similarly, it is in principle possible to design CCN codes by using generative adversarial networks and reinforcement learning methods for a model-free training framework. 

%%%%%%%%%%%%%%%%%%%%%%%%%%%%%%%%%%%%%%%%%%%%%%%%%%%%
\section{Conclusion}
We proposed a concatenation method to extend the code dimension of any neural encoder-decoder pair used for error correction such that practical code blocklengths and rates are achieved with significantly lower complexity than standard neural code constructions. Each codeword symbol of a non-binary outer classic block code is sequentially given as an input to the same inner neural encoder and, after transmitting the neural encoder outputs through a noisy channel, the remaining errors in the neural decoder outputs are corrected by the outer block decoder. This concatenation scheme allows to achieve exponentially-decreasing BLERs with low complexity. We illustrated that a CCN code with a high-rate outer RS code and an inner neural code under one-hot encoding significantly improves the BLER performance over the neural code for an AWGN channel. CCN codes are also robust to channel model changes, and we listed various extensions to improve the BLER performance of CCN codes. The full error-correction potential of the proposed code construction can be realized by implementing the listed extensions. Our results aimed to promote further research into concatenating classic and neural codes, which will help in making neural codes practical. 
%We also consider other code concatenation methods, such as parallel concatenation, that consist of classic and neural codes as CCN codes, which might be also useful to improve performance.
%In future work, we will analyze effects of the resolution of neural encoder outputs on the BLER performance of CCN codes.

\section*{Acknowledgment}
   This work was supported by the German Federal Ministry of Education and Research (BMBF) under the Grants 16KIS1004 and 16KIS1242, by the ELLIIT funding endowed by the Swedish government, and by the ZENITH Research and Career Development Award.

%\IEEEtriggeratref{0}
\vfill
\bibliographystyle{IEEEtran}
\bibliography{referencesCCNCodes}
\end{document}

%% file: Figure1.tex
\def\encpos{4.5}
\def\hidlsep{0.7}
\def\decpos{12.35}
\def\RS{-0.6}
\def\OH{1.3}

		\begin{tikzpicture}
		\tikzstyle{hidden neuron}=[fill,circle,inner sep=1.5pt];
		\tikzstyle{inner dot}=[fill,circle,inner sep=0.6pt];
		
		\coordinate (p11) at (\RS,0);
		\coordinate (p12) at (\RS,0);
		\node[block, very thick, minimum width=1.2cm, minimum height=3cm, align=center, fill=MyGray1!40, draw] [fit = (p11) (p12)] (B) {};
		\node[align=center, rotate=270] at (B.center) {Batch of\\RS sequences ${\tilde{u}}^{k_1}$};
		
		\coordinate (p21) at (\OH,0);
		\coordinate (p22) at (\OH,0);
		\node[block, very thick, minimum width=1.5cm, minimum height=3cm, align=center, fill=MyGray1!40, draw] [fit = (p21) (p22)] (rshp) {};
		\node[align=center, rotate=270] at (rshp.center) {One-hot encoding\\ of each sequence \\to $\mathbf{s}\in \mathbb{F}_2^{2^{k_1}}$};

		%%%%%%%%%% Encoder %%%%%%%%%%%
		
		\node at (\encpos,2.6) (nE) {neural encoder};
		\coordinate (p31) at (\encpos,0);
		\coordinate (p32) at (\encpos,0);
		\node[block, very thick, minimum width=4cm, minimum height=4.7cm, align=center, fill=MyGreen1!25, draw] [fit = (p31) (p32)] (E) {};
		%\node[align=center,font=\large] at (E1.center) {Enc.1 \\ $(k_1,n_1)$\\code};
		\node[align=center] at (E.center) {};
		
		% draw input nodes
		\foreach \x in {1,2,3,4,5,6}
            \node[hidden neuron] (c1\x) at (\encpos-\hidlsep*2,1.4-\x*0.4) {};
		\draw [-, densely dotted, thick] (c11)--(c16);
		
		% draw hidden nodes layer 1
		\foreach \x in {0,1,2,3,4,5,6,7}
            \node[hidden neuron] (c2\x) at (\encpos-\hidlsep,1.4-\x*0.4) {};
		\draw [-, densely dotted, thick] (c20)--(c27);
		
		% draw hidden nodes middle
		\foreach \x in {0,1,2,3,4,5,6,7}
            \node[hidden neuron] (cm\x) at (\encpos,1.4-\x*0.4) {};
		\draw [-, densely dotted, thick] (cm0)--(cm7);
		
		% dots in the middle
		\node[inner dot] (d0) at (\encpos-.5,0){};
		\node[inner dot] (d1) at (\encpos-.35,0){};
		\node[inner dot] (d3) at (\encpos-.2,0){};
		
		\node[inner dot] (d0) at (\encpos+.5,0){};
		\node[inner dot] (d1) at (\encpos+.35,0){};
		\node[inner dot] (d3) at (\encpos+.2,0){};
		
		% draw hidden nodes layer 3
		\foreach \x in {0,1,2,3,4,5,6,7}
            \node[hidden neuron] (c3\x) at (\encpos+\hidlsep,1.4-\x*0.4) {};
		\draw [-, densely dotted, thick] (c30)--(c37);

		% draw output nodes
		\foreach \x in {1,2,3,4,5,6}
            \node[hidden neuron] (c4\x) at (\encpos+\hidlsep*2,1.4-\x*0.4) {};
		\draw [-, densely dotted, thick] (c41)--(c46);
		
		% connect all dots
		\foreach \x in {0,1,2,3,4,5,6,7}
		{
		    \draw [-, ] (c11)--(c2\x);
		    \draw [-, ] (c16)--(c2\x);
		    \draw [-, ] (c14)--(c2\x);
		    \draw [-, ] (c3\x)--(c41);
		    \draw [-, ] (c3\x)--(c43);
		    \draw [-, ] (c3\x)--(c46);
        }
		
		\node[align=center,font=\footnotesize] at (\encpos-\hidlsep*2,1.6) {input\\size\\$2^{k_1}$};
		\node[align=center,font=\footnotesize] at (\encpos+\hidlsep*2,1.6) {output\\size\\$n_1$};
		
		\node (p1) at (\encpos-\hidlsep,-1.38){};
		\node (p2) at (\encpos+\hidlsep,-1.38){};
		%\draw [BC] (p1)--(p2);
		\node[align=center,font=\footnotesize] at (\encpos,-1.98) {$L$ hidden layers of size $2^{k_1}$};

		%%%%%%%%%% Decoder %%%%%%%%%%%
		% Begin
		
		\node at (\decpos,2.6) (nD) {neural decoder};
		\coordinate (p31) at (\decpos,0);
		\coordinate (p32) at (\decpos,0);
		\node[block, very thick, minimum width=4cm, minimum height=4.7cm, align=center, fill=MyGreen1!25, draw] [fit = (p31) (p32)] (D) {};
		\node[align=center] at (D.center) {};
		
		% draw input nodes
		\foreach \x in {1,2,3,4,5,6}
            \node[hidden neuron] (cc1\x) at (\decpos-\hidlsep*2,1.4-\x*0.4) {};
		\draw [-, densely dotted, thick] (cc11)--(cc16);
		
		% draw hidden nodes layer 1
		\foreach \x in {0,1,2,3,4,5,6,7}
            \node[hidden neuron] (cc2\x) at (\decpos-\hidlsep,1.4-\x*0.4) {};
		\draw [-, densely dotted, thick] (cc20)--(cc27);
		
		% draw hidden nodes middle
		\foreach \x in {0,1,2,3,4,5,6,7}
            \node[hidden neuron] (ccm\x) at (\decpos,1.4-\x*0.4) {};
		\draw [-, densely dotted, thick] (ccm0)--(ccm7);
		
		% dots in the middle
		\node[inner dot] (dd0) at (\decpos-.5,0){};
		\node[inner dot] (dd1) at (\decpos-.35,0){};
		\node[inner dot] (dd3) at (\decpos-.2,0){};
		
		\node[inner dot] (dd0) at (\decpos+.5,0){};
		\node[inner dot] (dd1) at (\decpos+.35,0){};
		\node[inner dot] (dd3) at (\decpos+.2,0){};
		
		% draw hidden nodes layer 3
		\foreach \x in {0,1,2,3,4,5,6,7}
            \node[hidden neuron] (cc3\x) at (\decpos+\hidlsep,1.4-\x*0.4) {};
		\draw [-, densely dotted, thick] (cc30)--(cc37);
		
		% draw output nodes
		\foreach \x in {1,2,3,4,5,6}
            \node[hidden neuron] (cc4\x) at (\decpos+\hidlsep*2,1.4-\x*0.4) {};
		\draw [-, densely dotted, thick] (cc41)--(cc46);
		
		% connect all dots
		
		\foreach \x in {0,1,2,3,4,5,6,7}
		{
		    \draw [-, ] (cc11)--(cc2\x);
		    \draw [-, ] (cc16)--(cc2\x);
		    \draw [-, ] (cc14)--(cc2\x);
		    \draw [-, ] (cc3\x)--(cc41);
		    \draw [-, ] (cc3\x)--(cc43);
		    \draw [-, ] (cc3\x)--(cc46);
        }
		
		\node[align=center,font=\footnotesize] at (\decpos-2*\hidlsep,1.6) {input\\size\\$n_1$};
		\node[align=center,font=\footnotesize] at (\decpos+2*\hidlsep,1.6) {output\\size\\$2^{k_1}$};
		\node[align=center,font=\footnotesize] at (\decpos,-1.98) {$L$ hidden layers of size $2^{k_1}$};
		
		% End
		%%%%%%%%%%%%%%%%%%%%%

		\coordinate (p21) at (7.65,0);
		\coordinate (p22) at (7.65,0);
		\node[block, very thick, minimum width=1.2cm, minimum height=3cm, align=center, fill=MyGray1!40, draw] [fit = (p21) (p22)] (pwr) {};
		\node[align=center, rotate=270] at (pwr.center) {Power\\ normalization};
		
		\coordinate (p31) at (9.25,0);
		\coordinate (p32) at (9.25,0);
		\node[block, very thick, minimum width=1cm, minimum height=3cm, align=center, fill=MyGray1!40, draw] [fit = (p31) (p32)] (Ch) {};
		\node[align=center,rotate=270] at (Ch.center) {Channel};
		
		\coordinate (p41) at (15.35,0);
		\coordinate (p42) at (15.35,0);
		\node[block, very thick, minimum width=1.2cm, minimum height=3cm, align=center, fill=MyGray1!40, draw] [fit = (p41) (p42)] (out) {};
		\node[align=center, rotate=270] at (out.center) {Softmax \\to $\hat{\mathbf{s}} \in (0,1)^{2^{k_1}}$};
		
		\coordinate (p41) at (15.35,0);
		\coordinate (p42) at (15.35,0);
		\node[block, very thick, minimum width=1.2cm, minimum height=3cm, align=center, fill=MyGray1!40, draw] [fit = (p41) (p42)] (sm) {};
		\node[align=center, rotate=270] at (sm.center) {Softmax \\to $\hat{\mathbf{s}} \in (0,1)^{2^{k_1}}$};
        
        \coordinate (p41) at (17,0);
		\coordinate (p42) at (17,0);
		\node[block, very thick, minimum width=1.2cm, minimum height=3cm, align=center, fill=MyGray1!40, draw] [fit = (p41) (p42)] (out) {};
		\node[align=center, rotate=270] at (out.center) {Arg max to $\hat{\tilde{u}}^{k_1}$};

		\draw [->,very thick] (B)--(rshp);
		\draw [->,very thick] (rshp)--(E);
		\draw [->,very thick] (E)--(pwr);
		\draw [->,very thick] (pwr)--(Ch);
		\draw [->,very thick] (Ch)--(D);
		\draw [->,very thick] (D)--(sm);
		\draw [->,very thick] (sm)--(out);
		
		\end{tikzpicture}